\documentclass[preprint,12pt]{elsarticle}  
\usepackage{graphicx} 
\usepackage{amssymb} 
\usepackage{amsmath} 

\begin{document} 

\begin{frontmatter} 

\title{Meson Assisted Strange Dibaryons} 

\author{A.~Gal} 

\address{Racah Institute of Physics, The Hebrew University, 91904 
Jerusalem, Israel} 

\begin{abstract} 

The state of the art in dibaryons with strangeness is reviewed, including 
the $K^-pp$ dibaryon which signals the onset of $\bar K$--nuclear binding. 
A new type of strange dibaryons is highlighted, where the primary binding 
mechanism is provided by strong $p$-wave pion interactions, as demonstrated 
by a quasibound $(I=\frac{3}{2},J^P=2^+)$ $\pi YN$ dibaryon calculation.

\end{abstract} 

\end{frontmatter} 

\section{Introduction} 
\label{sec:intro} 

The quark model (QM) has been very successful in reproducing SU(3) flavor 
octet (${\bf 8}$) and decuplet (${\bf 10}$) baryon masses, and to 
a somewhat lesser extent also higher mass baryon resonances, in terms of 
three-quark ($3q$) configurations. It is remarkable then that subsequent 
predictions of $6q$ configurations of bound or quasibound dibaryons have not 
been found to be realized in Nature, and that todate there is not even one 
single unambiguously established dibaryon. This statement is challenged 
by recent indications from $np\to d\pi\pi$ reactions at CELSIUS-WASA of 
a resonance structure at $M_R \approx 2.37$ GeV and $\Gamma_R \approx 70$ MeV 
that could be interpreted as a $\Delta\Delta$ dibaryon bound by about 100 MeV, 
but still about 200 MeV above the $d\pi\pi$ threshold \cite{BBB09}. We note 
that quark cluster calculations for $L=0$ $6q$ configurations \cite{OYa80} 
come up with only a weakly bound $(I,J^P)=(0,3^+)$ $\Delta\Delta$ dibaryon 
which in terms of a $NN$ configuration corresponds to a high-lying $^3D_3$ 
$pn$ resonance. 

In the strange sector, Jaffe's dibaryon $H$ with strangeness ${\cal S}=-2$ and 
quantum numbers $(I,J^P)=(0,0^+)$ which was predicted as a genuinely bound 
state well below the $\Lambda\Lambda$ threshold, perhaps the most cited 
ever prediction made for any dibaryon \cite{Jaf77}, has not been confirmed 
experimentally in spite of several comprehensive searches \cite{Bas97}. 
Another equally ambitious early prediction was made by Goldman {\it et al.} 
\cite{Gold87}, also using a variant of the MIT bag model, for ${\cal S}=-3$ 
dibaryons with $(I,J^P)$ values $(\frac{1}{2},1^+)$ and $(\frac{1}{2},2^+)$, 
dominated by $\Omega N$ structure and lying below the $\Xi\Lambda$ threshold. 
More realistic quark cluster calculations by Oka {\it et al.} \cite{OSI83}, 
applying resonating group methods, did not confirm Jaffe's deeply bound 
$H$, placing it just below the $\Xi N$ threshold as a resonance about 26 MeV 
above the $\Lambda\Lambda$ threshold. The underlying binding mechanism common 
to all of these orbital angular momentum $L=0$ configurations is the 
color-magnetic gluon exchange interaction between quarks, a feature 
emphasized by Oka \cite{Oka88} who systematically studied $L=0$ dibaryon 
configurations that could benefit from short-range attraction. Following 
earlier quark cluster calculations \cite{OYa80,OSI83}, these calculations 
resulted in no strange dibaryon bound states, and for the $\Omega N$-dominated 
${\cal S}=-3$ bound-state configurations predicted in Ref.~\cite{Gold87}, 
in particular, only a $(I,J^P)=(\frac{1}{2},2^+)$ quasibound state resulted. 

\begin{figure}[htb] 
\begin{center} 
\includegraphics[width=0.7\textwidth]{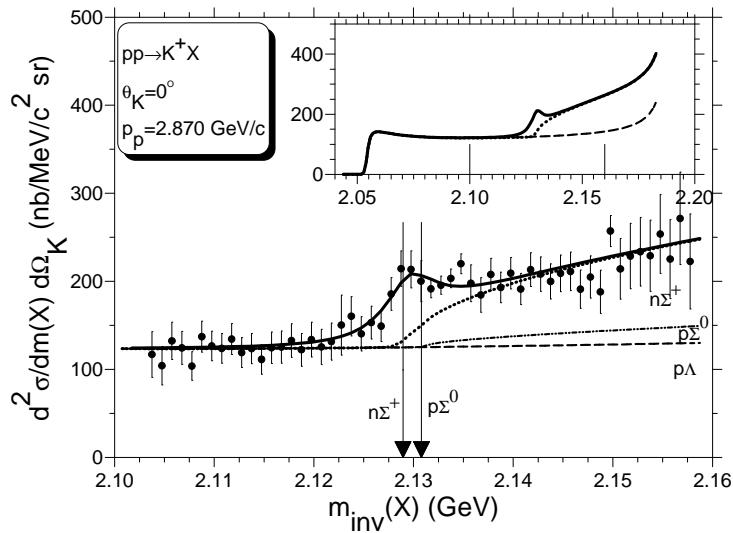}  
\caption{$YN$ invariant mass spectrum taken by the HIRES Colaboration at 
COSY \cite{cusp10} around the $\Sigma N$ threshold. The solid curve is 
a combined fit to $pp \rightarrow K^+(\Lambda p,\Sigma^0 p,\Sigma^+ n)$, 
including a BW fit to the threshold cusp.} 
\label{fig:cusp} 
\end{center} 
\end{figure} 

\begin{figure}[htb] 
\begin{center} 
\includegraphics[width=0.6\textwidth]{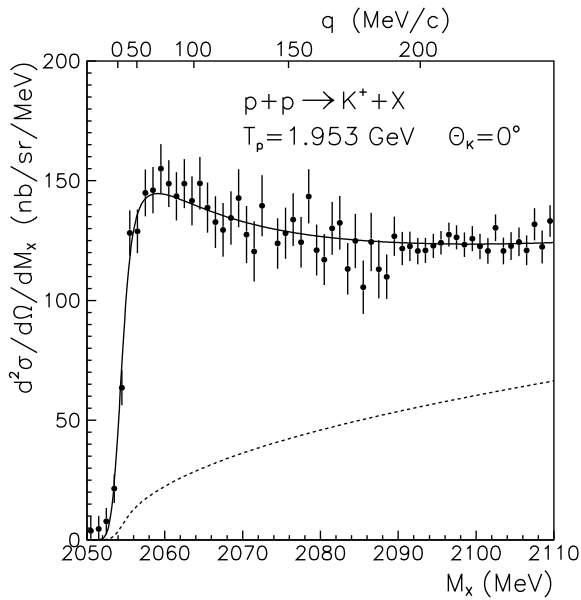} 
\caption{$\Lambda N$ invariant mass spectrum below the $\Sigma N$ threshold 
taken by the HIRES Collaboration at COSY \cite{HIRES10}.}
\label{fig:HIRES} 
\end{center} 
\end{figure}

For strangeness ${\cal S}=-1$, the focus of this review, old 
$K^-d\to\pi^-\Lambda p$ data \cite{Tan69} suggested resonant $\Lambda p$ 
structures at the $\Sigma N$ threshold and 10 MeV above it. However, 
a $(I,J^P)=(\frac{1}{2},1^+)$ $\Sigma N$ quasibound state is not necessarily 
required in order to reproduce the general shape of the $\Lambda p$ spectrum, 
as shown by multichannel Faddeev calculations \cite{TGE79,TDD86}. A cusp-like 
structure at the $\Sigma N$ threshold region could arise from the particularly 
strong one pion exchange tensor interaction in the $\Lambda N -\Sigma N$ 
coupled-channel $^3S_1-{^3D}_1$ partial waves. Several low-lying $L=1$ 
$\Lambda N$ resonances were predicted in singlet and triplet configurations 
in a QM study by Mulders {\it et al.} \cite{MAS80}, but negative results, 
particularly for the singlet resonance, were reported in dedicated $K^-$ 
initiated experiments \cite{JHK92,CDG92} near the $\Sigma N$ threshold. 
$YN$ invariant mass spectra taken in recent $pp \rightarrow K^+ X$ experiments 
at COSY \cite{cusp10}, shown in Fig.~\ref{fig:cusp}, give evidence for a cusp 
behavior at the $\Sigma N$ threshold, but no further imminent structure below 
or above. The portion of the spectrum below the kinematical range spanned in 
Fig.~\ref{fig:cusp} is shown in detail in Fig.~\ref{fig:HIRES}, countering 
earlier evidence from Saturne \cite{Siebert94} for a dibaryon signal at 
$M_X=2097$ MeV.    

In the present paper, following a brief summary of dibaryon theoretical 
expectations, I introduce the notion of pion assisted dibaryons, 
$\pi BB^\prime$. The idea is to enhance the binding of a $L=0$ $BB^\prime$ 
configuration through the strong attraction provided by $p$-wave $\pi B$ 
resonances. For the $\pi NN$ system this scenario requires quantum numbers 
$(I=2,J^P=2^+)$ to maximize the attraction from each one of the $(3,3)$ 
resonating $\pi N$ subsystems, but then the $NN$ subsystem must have 
$(I_{NN}=1,J^P=1^+)$ which is Pauli forbidden. Hence this mechanism fails 
for ${\cal S}=0$, although $(I=2,J^P=2^+)$ is allowed 
for a $\pi NN$ configuration with $(I_{NN}=1,{^{2s+1}L}={^3P})$ and 
$\ell_{\pi}=0,2,4$, and also for a $\Delta N$ configuration, particularly in 
$L=0$. Dibaryons of this sort that cannot be composed of two nucleons were 
discussed long ago in the QM and termed `extraneous dibaryons' \cite{MAS78}. 
The lowest ones predicted are $(I=0,J^P=0^-,2^-)$ at $M \approx 2120$ MeV, 
with the $0^-$ subsequently assigned to a $\pi NN$ resonance suggested by 
observed anomalies in $(\pi^+,\pi^-)$ reactions on nuclei at $T_{\pi}=50$ 
MeV ($M_{\pi NN} \sim 2065$ MeV) \cite{BCS93}. Such a state can be realized 
for two nucleons in $(I_{NN}=1,{^{2s+1}L_J}={^1S_0})$ plus an $s$-wave pion. 
There has been no solid experimental support for this dibaryon candidate 
ever since. 

The limitations imposed by the Pauli principle in the ${\cal S}=0$ sector have 
no counterpart in the ${\cal S}=-1$ sector, where a $\pi\Lambda N$ stretched 
configuration $(I=\frac{3}{2},J^P=2^+)$ is allowed. A possible quasibound 
state in this configuration has been recently studied \cite{gg08,gg10} by 
solving three-body Faddeev equations with $^3S_1-{^3D}_1,~\Lambda N -\Sigma N$ 
coupled channels chiral QM local interactions, and coupled $\pi Y$ 
($Y\equiv\Lambda,\Sigma$) and $\pi N$ separable $p$-wave interactions fitted 
to the position and decay parameters of the $\Sigma(1385)$ and $\Delta(1232)$ 
resonances, respectively. The results exhibit strong sensitivity to the 
$p$-wave $\pi Y$ interaction, the least phenomenologically constrained 
interaction in this calculation, with a $\pi\Lambda N$ quasibound state 
persisting over a wide range of acceptable parametrizations. 

Finally, we briefly review the ongoing study, both theoretically and 
experimentally, of a more familiar example of meson assisted dibaryons: 
a $K^-pp$ quasibound state in the ${\cal S}=-1$ sector, driven by the 
$\Lambda(1405)$ which within $\bar K N - \pi\Sigma$ coupled channels 
appears as a $s$-wave $K^-p$ quasibound state (QBS). For a recent overview 
of $K^-pp$ and its implications to $\bar K$--nuclear QBS phenomenology, see 
Ref.~\cite{weise10},

\section{Deuteron-like dibaryon candidates} 
\label{sec:BB}

\begin{table}[hbt] 
\begin{center} 
\caption{\( {\cal S} \)$=-2,-3,-4$ deuteron-like $L=0$ dibaryon candidates 
from the Nijmegen meson exchange model NSC97 \cite{Rijken99} and from EFT 
predictions \cite{Haidenbauer10}. Plus means yes, minus means no.} 
\label{tab:EFT} 
\begin{tabular}{cccccc} 
\hline 
&&&&& \\ 
\multicolumn{1}{c}{Strangeness} & \multicolumn{1}{c}{\( {\cal S} \)$=-2$} & 
\multicolumn{3}{c}{\( {\cal S} \)$=-3$} & 
\multicolumn{1}{c}{\( {\cal S} \)$=-4$} \\
&&&&& \\ 
\multicolumn{1}{c}{$BB^\prime$} & \multicolumn{1}{c}{$\Sigma\Sigma$} & 
\multicolumn{1}{c}{$\Lambda\Xi$} & \multicolumn{2}{c}{$\Sigma\Xi$} & 
\multicolumn{1}{c}{$\Xi\Xi$} \\  
&&&&& \\
$(I,{^{2s+1}[L=0]_j})$ & $(2,{^{1}S_{0}})$ & $(\frac{1}{2},{^{1}S_{0}})$ & 
$(\frac{3}{2},{^{1}S_{0}})$&$(\frac{3}{2},{^{3}S_{1}})$&$(1,{^{1}S_{0}})$ \\ 
&&&&& \\
\hline 
&&&&& \\ 
NSC97 & $+$ & $-$ & $+$ & $+$ & $+$  \\ 
&&&&& \\ 
EFT   & $-$ & $+$ & $+$ & $-$ & $+$  \\  
&&&&& \\ 
\hline 
\end{tabular} 
\end{center} 
\end{table} 

Here we review $BB^\prime$ deuteron-like dibaryon candidate configurations 
made out of ${\bf 8}_{\rm f}$ baryons in which no explicit quark degrees of 
freedom are considered. A well known example is the ${\cal S}=0$ 
$(I=0,J^P=1^+)$ $NN$ weakly bound deuteron. It is established experimentally 
that for ${\cal S}=-1$, the $\Lambda N$ and $\Sigma N$ interactions are too 
weak to provide binding. The associated scattering data have been used in 
several $YN$ potential fits, respecting SU(3)$_{\rm f}$ within well defined 
symmetry breaking schemes, to make predictions for stranger $BB^\prime$ 
systems. There is almost general consensus, supported also by scarce 
$\Lambda\Lambda$ data, that there are no bound states for ${\cal S}=-2$, 
except for a $\Sigma\Sigma$ $(I=2,J^P=0^+)$ quasibound state. However, for 
${\cal S}=-3,-4$ there are several bound-state candidates, as listed in 
Table~\ref{tab:EFT}. The bound states listed in the table are due to two 
methodologies: 
\begin{itemize}  
\item  
The latest Nijmegen extended-soft-core (ESC) meson-exchange model, with broken 
SU(3)$_{\rm f}$ for ${\cal S}=0,-1,-2$, claims no quasibound states (QBS) 
\cite{Rijken10} while providing no predictions yet for stranger systems. 
Earlier soft-core versions, e.g. NSC97 \cite{Rijken99}, found QBS for 
${\cal S}=-3,-4$ as listed in Table~\ref{tab:EFT}. The three $^{1}S_{0}$ bound 
states in this model are in one-to-one correspondence with $BB^\prime$ states 
assigned in SU(3)$_{\rm f}$ to the ${\bf 27}_{\rm f}$ representation which 
includes the $NN$ $^{1}S_{0}$ virtual state. Similarly, the $^{3}S_{1}$ bound 
state in this model, as listed in the table, is also the only $BB^\prime$ 
state which together with the deuteron is assigned in SU(3)$_{\rm f}$ to 
the ${\overline {\bf 10}}_{\rm f}$ representation. 
\item 
The Bonn-J\"{u}lich chiral effective field theory (EFT) model, applied 
in lowest order to ${\cal S}=-1,-2,-3,-4$ with low energy constants 
(LEC) constrained by SU(3)$_{\rm f}$ and fitted to low-energy $YN$ 
data \cite{Haidenbauer06} finds no QBS for ${\cal S}=-1$ and also for 
${\cal S}=-2$ \cite{Haidenbauer07}. The ${\cal S}=-3,-4$ sectors require 
only the five LECs determined in the $YN$ sector fit, independently of the 
sixth LEC required in the ${\cal S}=-2$ sector (this LEC is consistent 
with zero). This is how one gets predictions \cite{Haidenbauer10} for 
the ${\cal S}=-3,-4$ dibaryon candidates listed in Table~\ref{tab:EFT}. 
The predicted binding energies are all in the several MeV range, 
in agreement with what one expects for deuteron-like dibaryons where 
pion dynamics is dominant. 
The model dependence of these predictions is assessed within the model 
by varying a cutoff momentum in the range $550-700$ MeV/c. Additional 
model dependence is likely to arise in next-to-leading-order evaluations. 
While the predictions of the EFT model differ for some states and agree 
for other ones with those of the NSC97 model, both these models predict 
strong attraction for all the configurations listed in the table. 
\end{itemize}

\section{Six-quark dibaryon configurations} 
\label{sec:6q} 

Early discussions of $6q$ dibaryons were based on symmetry considerations 
related to the assumed dominance of a color-magnetic (CM) gluon exchange 
interaction 
\begin{equation} 
V_{CM}=\sum_{i<j}-(\lambda_i\cdot\lambda_j)(s_i\cdot s_j)v(r_{ij}), 
\label{eq:cm} 
\end{equation} 
where $\lambda_i$ and $s_i$ are the color and spin operators of the $i$-th 
quark and $v(r_{ij})$ is a flavor-conserving short-ranged spatial interaction 
between quarks $i,j$. For $L=0$ orbitally symmetric color-singlet 
$n$-quark cluster, the matrix element of $v(r_{ij})$ is independent of the 
particular $i,j$ pair and is denoted ${\cal M}_0$, allowing for a closed 
form summation over $i$ and $j$ in Eq.~(\ref{eq:cm}) with the result: 
\begin{equation} 
\langle V_{\rm CM} \rangle = [-\frac{n(10-n)}{4}+\Delta{\cal P}_{\rm f}+
\frac{s(s+1)}{3}]{\cal{M}}_0, 
\label{eq:CM} 
\end{equation} 
where ${\cal P}_{\rm f}$ sums over $\pm 1$ for any symmetric/antisymmetric 
flavor pair, $\Delta{\cal P}_{\rm f}$ means with respect to the 
SU(3)$_{\rm f}$ $\bf 1$ antisymmetric representation of $n$ quarks, $n=3$ 
for baryons and $n=6$ for dibaryons, $s$ is the total Pauli spin, and where 
${\cal M}_0\sim 75$~MeV from the $\Delta - N$ mass difference. The leading 
${\cal S}=0,-1,-2,-3$ dibaryon candidates are listed in Table~\ref{tab:oka} 
following Ref.~\cite{Oka88}, where $\Delta \langle V_{\rm CM} \rangle$ 
stands for the CM interaction gain $\langle V_{\rm CM} {\rangle}_{6q} - 
\langle V_{\rm CM}{\rangle}_{B} -\langle V_{\rm CM} {\rangle}_{B^\prime}$ in 
the $6q$ dibaryon configuration with respect to the sum of CM contributions 
from the separate $B$ and $B^{\prime}$ $3q$ baryons that define the lowest 
$BB^{\prime}$ threshold. The table shows clearly the prominence of the 
${\cal S}=-2$ $H$ dibaryon. 
\begin{table}[hbt] 
\begin{center} 
\caption{Leading $6q$ $L=0$ dibaryon candidates \cite{Oka88}, their 
$BB^\prime$ structure and the CM interaction gain with respect of the lowest 
$BB^\prime$ threshold calculated by means of Eq.~(\ref{eq:CM}). Asterisks are 
used for the ${\bf 10}_{\rm f}$ baryons $\Sigma^\ast \equiv \Sigma(1385)$ and 
$\Xi^\ast \equiv \Xi(1530)$. The symbol [i,j,k] stands for the Young tablaux 
of the SU(3)$_{\rm f}$ representation, with i arrays in the first row, 
j arrays in the second row and k arrays in the third row, from which 
${\cal P}_{\rm f}$ is evaluated.} 
\label{tab:oka} 
\begin{tabular}{cccccc} 
\hline 
&&&&& \\ 
\( -{\cal S} \)& SU(3)$_{\rm f}$ & $I$ & $J^{\pi}$ & $BB^\prime$ structure & 
$\frac{\Delta \langle V_{\rm CM} \rangle}{M_0}$  \\  
&&&&& \\ 
\hline 
&&&&& \\ 
0 & [3,3,0] $\overline{\bf 10}$ & 0 & 3$^+$ & $\Delta\Delta$ & 0  \\ 
&&&&& \\ 
1 & [3,2,1] $\bf 8$ & 1/2 & 2$^+$ & 
$\frac{1}{\sqrt{5}}(N\Xi^{\ast}+2\Delta\Sigma)$ & 
$-1$  \\
&&&&& \\ 
2 & [2,2,2] $\bf 1$ & 0 & 0$^+$ & 
$\frac{1}{\sqrt{8}}(\Lambda\Lambda+2N\Xi-\sqrt{3}\Sigma\Sigma)$ & $-2$  \\
&&&&& \\ 
3 & [3,2,1] $\bf 8$ & 1/2 & 2$^+$ & $\frac{1}{\sqrt{5}}
(\sqrt{2}N\Omega-\Lambda\Xi^{\ast}+\Sigma^{\ast}\Xi-\Sigma\Xi^{\ast})$&$-1$ \\ 
&&&&& \\ 
\hline 
\end{tabular} 
\end{center} 
\end{table} 

\begin{figure}[htb] 
\begin{center} 
\includegraphics[width=0.55\textwidth]{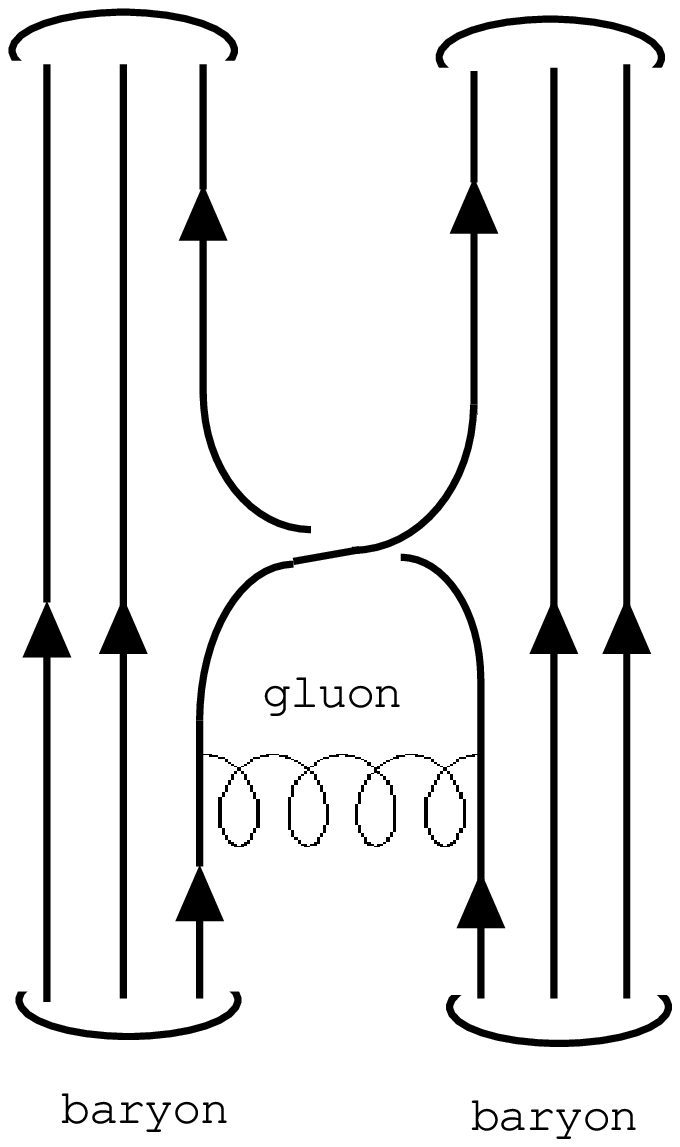} 
\includegraphics[width=0.44\textwidth]{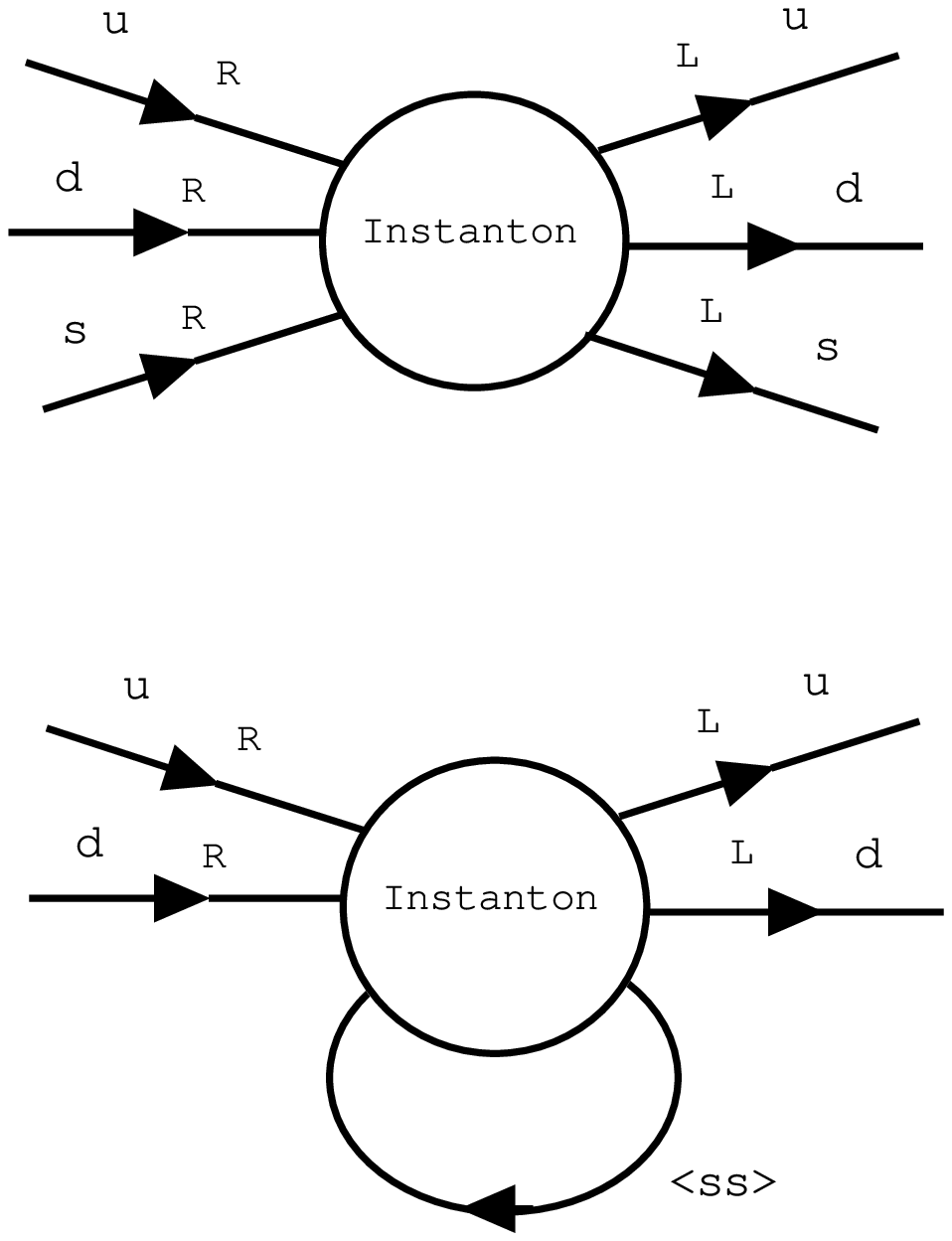} 
\caption{Quark interaction diagrams for $6q$ dibaryon calculations 
\cite{Oka88,Oka95}. Left: one gluon exchange accompanied by quark exchange. 
Right: instanton contributions.}
\label{fig:Oka} 
\end{center} 
\end{figure} 

More realistic $6q$ calculations, e.g. Refs.~\cite{Oka88,Oka91}, employ 
quark-cluster models (QCM) that break SU(3)$_{\rm f}$ and account for full 
quark antisymmetrization, also making contact via resonating group methods 
(RGM) with related $BB^\prime$ coupled channels and thresholds. Input 
interactions are shown in Fig.~\ref{fig:Oka}, taken from Ref.~\cite{Oka95}, 
where the left-hand side diagram corresponds to a gluon-exchange mediated 
$BB^\prime$ quark-exchange interaction and the right-hand side diagrams 
correspond to instanton induced interactions \cite{Koch85}, a $3q$ as well 
as a nonstrange $2q$ version. Recent RGM quark calculations by Fujiwara and 
collaborators \cite{Fujiwara07} do not use the instanton interactions, 
but add several meson exchanges to the gluon exchange interaction. None of 
these models produces a stable $6q$ dibaryon. This may be demonstrated for 
the $H$ within several frameworks, including the QCM, as follows: 
\begin{itemize} 
\item 
In the QCM, owing to the repulsive instanton induced interaction, the $H$ 
becomes barely bound or unbound, perhaps a $\Lambda\Lambda$ resonance below 
the $N\Xi$ threshold \cite{Oka91}. 
\item  
In QCD sum rule calculations, where $m_{H}$ is correlated with $m_{nn}$, 
the $H$ might be bound, although with a large uncertainty, $B_{H}=40\pm 70$ 
MeV \cite{Oka94}. 
\item 
Particle stability of $_{\Lambda\Lambda}^{~~6}{\rm He}$ rules out 
$B_{H}>7$~MeV, otherwise $_{\Lambda\Lambda}^{~~6}{\rm He}$ would have 
disintegrated into ${H}+{{^4}{\rm He}}$, at odds with its observed weak
decay. 
\end{itemize}

\section{Pion assisted dibaryons} 
\label{sec:pion} 

Here we sketch the recent calculation \cite{gg10} of a $(I,J^P)=(\frac{3}{2},
2^+)$ $\pi\Lambda N$ quasibound state, driven by the two-body $\pi N$ 
resonance $\Delta(1232)$ and the $\pi\Lambda -\pi\Sigma$ resonance 
$\Sigma(1385)$, for a $\Lambda N -\Sigma N$ $^3S_1-{^3D_1}$ coupled channels 
configuration. The $YN$ interaction was taken from a chiral QM calculation, 
with parameters constrained by acceptable low-energy parameters \cite{GVF07}. 
The $\pi B$ interactions were taken in a rank-one separable form
\begin{equation} 
\langle p|V_{\pi B,\pi B^{\prime}}|p^\prime\rangle=
(\gamma_B\gamma_B^\prime)^{\frac{1}{2}}g_{\pi B}(p)g_{\pi B^\prime}(p^\prime)~. 
\end{equation} 
The $\pi N\Delta(1232)$ $p$-wave form factor $g_{\pi N}$ was obtained from 
a very good fit of the $P_{33}$ phase shift to a functional form with three 
well constrained parameters (in addition to $\gamma_{\pi N}$): 
\begin{equation}  
g_{\pi N}(p)=p[e^{-p^2/\beta^2}+Ap^2e^{-p^2/\alpha^2}]~, 
\label{eq:piN}
\end{equation} 
\begin{equation}  
A=0.2~{\rm fm}^2,~\beta=1.31~{\rm fm}^{-1},~\alpha=3.21~{\rm fm}^{-1}. 
\label{eq:Aalphabeta} 
\end{equation}
The r.m.s. momentum associated with $g_{\pi N}(p)$ is 
${\langle p^2_{\pi N} \rangle}^{\frac{1}{2}}=5.55~{\rm fm}^{-1}=1095$ MeV/c, 
where 
\begin{equation} 
{\langle p_{\pi N}^2 \rangle}~=~\frac{\int_0^\infty 
g_{\pi N}(p)~p^2~d^3p}{\int_0^\infty g_{\pi N}(p)~d^3p}. 
\label{eq:p_piN} 
\end{equation} 
This high-momentum value does not rule out a spatial size of order 1 fm for 
$\Delta(1232)$. Indeed, if ${\tilde g}_{\pi N}(r)$ is the Fourier transform 
of $g_{\pi N}(p)$, for $\ell=1$, then ${\langle r^2_{\pi N} \rangle}^
{\frac{1}{2}}=0.875$ fm. The $\pi Y \Sigma(1385)$ coupled-channel $p$-wave 
form factor was fitted to the position, width and decay branching ratios of 
$\Sigma(1385)$, using the form 
\begin{equation} 
g_{\pi Y}(p)=p(1+Ap^2)e^{-p^2/\alpha^2}~, 
\label{eq:piY} 
\end{equation} 
leaving room for gridding over one of four fit parameters which was chosen 
to be $A$. The sensitivity of the quasibound calculation to values of $A$ 
in the range $1.0-1.8~{\rm fm}^2$ is demonstrated by the last four columns 
in Table~\ref{tab:piY}. It is by far the strongest sensitivity in this 
calculation, remarkably so for a fairly small variation over the range of 
r.m.s. momenta associated with the $p$-wave form factor $g_{\pi Y}(p)$. 

\begin{table} 
\begin{center} 
\caption{$\pi\Lambda N$ binding energy (in MeV) calculated \cite{gg10} for 
four $\pi Y$ interaction models specified by ${\langle p^{2}_{\pi Y} \rangle}
^{\frac{1}{2}}$ and six chiral QM versions of the $^3S_1 - {^3}D_1$ $YN$ 
interaction fitted to given $\Lambda N$ scattering length $a$ and effective 
range $r_0$. The momentum $p_{\rm lab}(\delta=0)$ is the $\Lambda$ laboratory 
momentum where the $^3S_1$ $\Lambda N$ phase shift changes sign. The last row 
corresponds to switching off the $YN$ interaction.} 
\label{tab:piY} 
\begin{tabular}{ccccccc} 
\hline 
\multicolumn{3}{c}{$YN$ interaction} &
\multicolumn{4}{c}{${\langle p^{2}_{\pi Y} \rangle}^{\frac{1}{2}}$ 
(fm$^{-1}$)}  \\ 
$a$ & $r_0$ & $p_{\rm lab}(\delta=0)$ & 3.91 & 3.76 & 3.60 & 3.48 \\ 
(fm) & (fm) & (MeV/c) & \multicolumn{4}{c}{$B_{\pi\Lambda N}$ (MeV)} \\
\hline 
$-1.35$ &   3.39  &   987    &   99  & 65  & 30  & 6   \\ 
$-1.40$ &   3.32  &  1011    &   99  & 66  & 30  & 6   \\ 
$-1.64$ &   3.09  &  1146    &  102  & 68  & 32  & 8   \\ 
$-1.71$ &   3.03  &  1198    &  102  & 68  & 33  & 9   \\ 
$-1.78$ &   2.98  &  1272    &  103  & 69  & 33  & 9   \\ 
$-1.86$ &   2.93  &  1446    &  104  & 69  & 34  & 10  \\ 
 
   --   &   --    &    --    &  120  & 84  & 47  & 21  \\ 
\hline  
\end{tabular} 
\end{center} 
\end{table} 

Irrespective of which $\pi Y$ model is chosen, the $YN$ interaction 
always produces repulsion, thus lowering the calculated binding energy, 
as demonstrated in the last line of Table~\ref{tab:piY} which corresponds 
to switching off the $YN$ interaction. This repulsive $YN$ effect owes its 
origin to the high-momentum components of the $\pi B$ form factors which 
within the three-body calculation highlight the short-range repulsive 
region of the $YN$ interaction. All in all, the calculations described above 
leave sufficient room for a quasibound $S=-1$ dibaryon, here denoted $\cal D$, 
decaying to a $d$-wave $I=\frac{3}{2}$ $\Sigma N$ scattering state and perhaps 
also to $(\pi \Lambda N)_{I=\frac{3}{2}}$ if it corresponds to $\pi \Sigma N$ 
quasibound state above the $\pi \Lambda N$ threshold. To search for $\cal D$, 
the following reactions are possible: 
\begin{equation} 
K^- + d \to  {\cal D}^- + \pi^+~, \,\,\,\,\,\,\,\,\,\, 
\pi^- + d \to {\cal D}^- + K^+~,  
\label{eqKpiK}
\end{equation}   
\begin{equation} 
p + p \to {\cal D}^+ + K^+~.  
\label{eqpp} 
\end{equation}

\section{Kaon assisted dibaryons} 
\label{sec:kaon} 

The low-energy $\bar K N$ $s$-wave interaction is particularly strong in the 
$I=0$ channel, leading to a $\bar K N -\pi\Sigma$ QBS, the $\Lambda(1405)$, 
nominally 27 MeV below the $K^-p$ threshold and with a decay width of about 
50 MeV to the $\pi\Sigma$ channel. The lightest $\bar K NN$ dibaryon 
configuration maximizing the strongly attractive $I=0~\bar K N$ interaction 
is $[\bar K (NN)_{I=1}]_{I=1/2,J^{\pi}=0^-}$, loosely denoted as 
$K^-pp$. Several few-body calculations of $K^-pp$ are summarized in 
Table~\ref{tab:K^-pp}. 

\begin{table}[hbt] 
\begin{center} 
\caption{Calculated $K^-pp$ binding energies ($B_{K^-pp}$), mesonic 
($\Gamma_{\rm m}$) \& nonmesonic ($\Gamma_{\rm nm}$) widths (in MeV).} 
\begin{tabular}{ccccccc} 
\hline 
& \multicolumn{4}{c}{single ${\overline K}N - \pi\Sigma$ QBS pole} 
& \multicolumn{2}{c}{two poles} \\ 
& \multicolumn{2}{c}{variational} & \multicolumn{3}{c}{Faddeev} & 
\multicolumn{1}{c}{variational}  \\  
& \cite{YA02,AY02}&\cite{WG08} & \cite{SGM07}&\multicolumn{2}{c}{\cite{IS10}}& 
\cite{DHW08}  \\  \hline
$B_{K^-pp}$ & 48 & 40--80 & 50--70 & 44--58 & 9--16 & 17--23  \\ 
$\Gamma_{\rm m}$ & 61 & 40--85 & 90--110 & 34--40 & 34--46  & 40--70 \\ 
$\Gamma_{\rm nm}$ & 12 & $\sim 20$ & & & & 4--12  \\  \hline 
\end{tabular} 
\label{tab:K^-pp} 
\end{center} 
\end{table} 

\begin{figure}[thb] 
\begin{center} 
\includegraphics[width=0.5\textwidth]{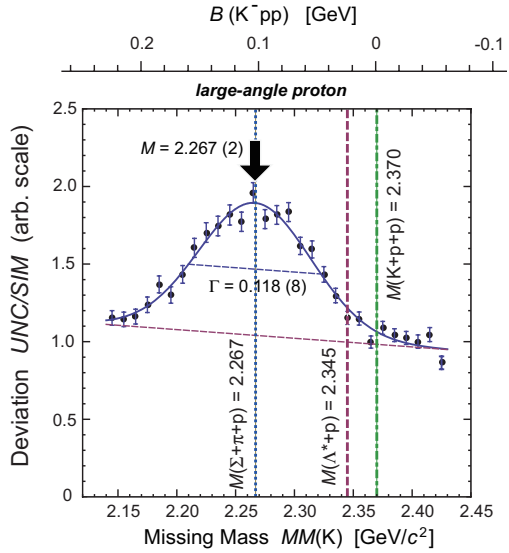} 
\caption{DISTO reanalysis of $K^+$ missing-mass spectrum in 
$pp\to K^+\Lambda p$ at $T_p=2.85$~GeV \cite{Yam10}.} 
\label{fig:disto} 
\end{center} 
\end{figure} 

The table supports the expectation that the $K^-pp$ system is bound, although 
there are marked differences between the values calculated for the binding 
energy $B_{K^-pp}$. The input to all of the listed calculations is constrained 
by requiring that the $\Lambda(1405)$ $\pi\Sigma$ resonance position, and as 
much as possible also its shape, are reproduced. The models that achieve it 
with one ${\bar K} N - \pi\Sigma$ QBS pole, necessarily at or in the immediate 
vicinity of $\sqrt{s}=1405$ MeV, produce $K^-pp$ binding in the range $40-80$ 
MeV, considerably higher than the $10-20$ MeV range in models with two 
${\bar K} N - \pi\Sigma$ QBS poles. Among the listed calculations, only 
Refs.~\cite{IS10,DHW08} used chiral ${\bar K} N - \pi\Sigma$ models that 
produce two such poles, but in the first listed Faddeev calculation of 
Ref.~\cite{IS10} the energy dependence of the ${\bar K} N - \pi\Sigma$ 
coupled channels input was suppressed in favor of fixed threshold values, 
thus making it effectively a single-pole calculation.{\footnote{The second 
listed Faddeev calculation of Ref.~\cite{IS10}, in addition to a relatively 
shallow $K^-pp$ QBS listed in the column before last, also produced a deeply 
lying and very broad $K^-pp$ QBS (unlisted in the table) in the range of 
$B_{K^-pp}=(67-89)$ MeV and $\Gamma_{\rm m}=(244-320)$ MeV.}} The shallow QBS 
of the last two columns are primarily related to the pole position of the 
chiral $\bar K N$ amplitude which resonates at 1420 MeV (close to the upper 
of the two ${\bar K} N - \pi\Sigma$ poles) in chiral models, whereas in 
single-pole models the $\bar K N$ amplitude resonates necessarily at 1405 MeV. 
This correlation with the resonance behavior of the $\bar K N$ amplitude 
has been verified in the variational calculations of Ref.~\cite{WG08} and in 
the coupled-channel Faddeev study of Ref.~\cite{SGMR07}. 
A notable feature of the $K^-pp$ single-pole coupled-channel calculations 
\cite{WG08,SGM07} in Table~\ref{tab:K^-pp} is that the explicit use of the 
$\pi\Sigma N$ channel adds about $20 \pm 5$~MeV to the $K^-pp$ binding energy 
with respect to that calculated using effective $\bar K N$ potential within 
a $\bar K NN$ single-channel calculation \cite{IS08}. 

A reanalyzed DISTO spectrum of $pp\to K^+\Lambda p$ from Ref.~\cite{Yam10} 
is shown in Fig.~\ref{fig:disto}. The authors claim that the peak structure 
of the outgoing $\Lambda p$ gives evidence for a $K^-pp$ quasibound state 
decaying via $K^-pp \to \Lambda p$. The location of the peak practically on 
top of the $\pi\Sigma N$ threshold, and its large width, are at odds with 
any of the few-body calculations listed in Table~\ref{tab:K^-pp}, posing 
a problem for a $K^-pp$ QBS interpretation. At present, besides ongoing 
$p(p,K^+)$ measurements at GSI to improve on the DISTO data, the $K^-pp$ 
system will be explored at J-PARC in the $^3{\rm He}(K^-,n)$ and 
$d(\pi^+,K^+)$ reactions.

\section{Summary and outlook} 
\label{sec:sum} 

In this Festschrift contribution I have reviewed the state of the art in 
dibaryons with strangeness, including $K^-pp$ for which extensive experimental 
searches are underway. A new class of strange dibaryons which are termed meson 
assisted strange dibaryons was highlighted. $K^-pp$ is just one example, where 
the strong $s$-wave $K^-p$ interactions stabilize the initially unstable $pp$ 
dibaryon. Pion assisted dibaryons $\pi B B^\prime$ offer more possibilities by 
making use of the strong $p$-wave $\pi B$ and $\pi B^\prime$ resonances 
classified in the SU(3) ${\bf 10}_{\rm f}$ representation and its 
extensions into charm ${\cal C}\neq 0$. Possible candidates are as follows 
\cite{gg08}. 
\begin{itemize} 
\item ${\cal S}\neq 0,~{\cal C}=0$: 
\begin{equation} 
{\cal S}=-1:~\pi\Lambda N,~~{\cal S}=-2:~\pi\Xi N,~~{\cal S}=-3:~
\pi\Lambda\Xi, 
\end{equation} 
\item ${\cal C}=+1$:  
\begin{equation}  
{\pi}N\Lambda_c(2286),~~~~{\pi}N\Xi_c(2470),~~~~{\pi}N\Omega_c(2700),
\end{equation} 
\begin{equation} 
\pi\Lambda\Lambda_c(2286),~~~~\pi\Lambda\Xi_c(2470),~~~~
\pi\Lambda\Omega_c(2700), 
\end{equation} 
\begin{equation} 
\pi\Xi\Lambda_c(2286),~~~~\pi\Xi\Xi_c(2470),~~~~\pi\Xi\Omega_c(2700). 
\end{equation} 
\item ${\cal C}=+2$: 
\begin{equation} 
\pi\Lambda_c(2286)\Xi_c(2470),~~\pi\Lambda_c(2286)\Omega_c(2700),~~
\pi\Xi_c(2470)\Omega_c(2700). 
\end{equation} 
\end{itemize} 
Note the appearance of the $\frac{1}{2}^+$ $\Omega_c$ baryon, of quark 
structure $ssc$. In the case of charmed baryons, the $p$-wave non-charmed 
SU(3)-${\bf 10}_{\rm f}$ $\frac{3}{2}^+$ resonances are replaced by charmed 
SU(3)-${\bf 6}_{\rm f}$ members of the SU(4)-${\bf 20}_{\rm f}$ extension: 
\begin{equation} 
\Sigma(1385)\to\Sigma_c(2520),~~~~\Xi(1530)\to\Xi_c(2645),~~~~
\Omega(1670)\to\Omega_c(2770). 
\end{equation} 
Here we limited listing to ${\cal C}=+1$ baryons. For a future charmed 
bound-state study, note that the $\pi N \Lambda_c(2286)$ threshold lies 
{\it below} $N \Sigma_c(2455)$, where $\Sigma_c(2455)$ is the lowest 
lying known $\Sigma_c$, with assumed $J^P=\frac{1}{2}^+$. Therefore, if 
$\pi N \Lambda_c(2286)$ is bound, it will decay only by weak interactions. 

Realistic calculations of pion assisted dibaryons have been reported so far 
only for the $\pi\Lambda N$ system with $(I=\frac{3}{2},J^P=2^+)$, where its 
coupling to the higher channel $\pi\Sigma N$ was considered while the coupling 
to the lower nonpionic channel $\Sigma N$ was ignored \cite{gg08,gg10}. 
This system is a good candidate for a quasibound dibaryon either below 
or above the $\pi\Lambda N$ threshold, but its precise location depends 
sensitively on the poorly known $\pi\Lambda\Sigma(1385)-\pi\Sigma\Sigma(1385)$ 
form factor. We note that the $\pi\Lambda N$ threshold is about 130 MeV below 
the $\Sigma(1385)N$ threshold and 230 MeV below the $\Sigma\Delta(1232)$ 
threshold, these latter thresholds being relevant in the discussion of the 
leading $6q$ ${\cal S}=-1$ dibaryon configuration of Table~\ref{tab:oka}. 
In this respect a $\pi\Lambda N$ dibaryon, if established, is the next one 
in excitation energy to the lowest ${\cal S}=-1$ thresholds of $\Lambda N$ 
and $\Sigma N$.

\section*{Acknowledgements} 
Gerry Brown has played a leading role in the development of nuclear physics 
and its derivatives. On this occasion of his 85th birthday Festschrift, I am 
pleased to acknowledge many inspiring suggestions he made to me throughout my 
own career. 
The work on pion assisted dibaryons is coauthored by Humberto Garcilazo who 
has contributed significantly to its development. I wish to also thank Wolfram 
Weise with whom I enjoyed several inspiring discussions on the $K^-pp$ 
dibaryon and its implications to $\bar K$--nuclear physics. This research is 
partially supported by the EU Initiative FP7, HadronPhysics2, under Project 
227431.

\end{document}